\documentclass[aps,prb,groupedaddress,notitlepage]{revtex4-1}

\usepackage{graphicx}
\usepackage{bm}

\begin{document}

%\preprint{}

\title{Ab initio theory of the spin-dependent conductivity tensor
       and the spin Hall effect in random alloys}


\author{I. Turek}
\email[]{turek@ipm.cz}
%\thanks{}
\affiliation{Institute of Physics of Materials,
Czech Academy of Sciences,
\v{Z}i\v{z}kova 22, CZ-616 62 Brno, Czech Republic}

\author{J. Kudrnovsk\'y}
\email[]{kudrnov@fzu.cz}
\affiliation{Institute of Physics, 
Czech Academy of Sciences,
Na Slovance 2, CZ-182 21 Praha 8, Czech Republic}

\author{V. Drchal}
\email[]{drchal@fzu.cz}
\affiliation{Institute of Physics, 
Czech Academy of Sciences,
Na Slovance 2, CZ-182 21 Praha 8, Czech Republic}


\date{\today}

\begin{abstract}

We present an extension of the relativistic electron transport theory
for the standard (charge) conductivity tensor of random alloys
within the tight-binding linear muffin-tin orbital method to the
so-called spin-dependent conductivity tensor, which describes the
Kubo linear response of spin currents to external electric fields.
The approach is based on effective charge- and spin-current operators,
that correspond to intersite electron transport and that are nonrandom,
which simplifies the configuration averaging by means of the coherent
potential approximation.
Special attention is paid to the Fermi sea term of the spin-dependent
conductivity tensor, which contains a nonzero incoherent part, in
contrast to the standard conductivity tensor.
The developed formalism is applied to the spin Hall effect in binary
random nonmagnetic alloys, both on a model level and for Pt-based
alloys with an fcc structure.
We show that the spin Hall conductivity consists of three
contributions (one intrinsic and two extrinsic) which exhibit
different concentration dependences in the dilute limit of an alloy.
Results for selected Pt alloys (Pt-Re, Pt-Ta) lead to the spin
Hall angles around 0.2; these sizable values are obtained for
compositions that belong to thermodynamically equilibrium phases.
These alloys can thus be considered as an alternative to other
systems for efficient charge to spin conversion, which are often
metastable crystalline or amorphous alloys. 

\end{abstract}

%\pacs{72.10.Bg, 72.15.Gd, 75.47.Np}
%\keywords{}

\maketitle

%\subsection{}
%\subsubsection{}

\section{Introduction\label{s_intr}}

Efficient generation and reliable detection of spin currents
in magnetoelectronic devices belong to the central topics of the
whole area of spintronics \cite{r_2012_mvs, r_2012_tz}.
In systems without local magnetic moments and in the absence of
external magnetic fields, the most important phenomenon in this
context is undoubtedly the spin Hall effect (SHE) \cite{r_2015_svw}. 
This transport phenomenon was predicted as a consequence of
spin-orbit interaction \cite{r_1999_jeh};
subsequent systematic theoretical and experimental investigation
resulted in detailed understanding of its basic aspects.

In pure metals and ordered crystals, the SHE arises solely due to
the band structure of the system, i.e., due to the dependence of
energy eigenvalues and eigenvectors on the reciprocal-space vector
$\mathbf{k}$, without the need of any explicit mechanism of
electron scattering.
The central quantity, namely the intrinsic spin Hall conductivity
(SHC), can be expressed in terms of the Berry curvature of the
occupied electron states \cite{r_2008_gmc, r_2010_fbm}.
In diluted metallic alloys, the impurity scattering leads to an
extrinsic contribution to the SHC that can be treated on different
levels of sophistication of electron transport theory.
The extrinsic SHC is due to the skew-scattering and side-jump
mechanisms whereby the former one dominates in the dilute limit. 
Evaluation of the extrinsic contribution
within the linearized Boltzmann equation rests on
the scattering-in terms \cite{r_2010_gfz} while the Kubo linear
response theory with the coherent potential approximation (CPA)
requires inclusion of the so-called vertex corrections
\cite{r_2011_lgk}.

The intrinsic and extrinsic contributions to the SHC have its
counterparts in a description of the anomalous Hall effect in
ferromagnetic systems \cite{r_2010_nso, r_2014_zck}, so that
similar concepts appear in studies of both transverse transport
phenomena.
In the SHE, the relative magnitude of the SHC is expressed by
the so-called spin Hall angle (SHA), defined as a ratio of the
SHC to the standard (charge) longitudinal conductivity.
The SHA represents a dimensionless measure of efficiency of
the charge to spin conversion.
For this reason, both experimental and theoretical effort has
recently been devoted to find systems with large SHA, see, e.g.,
Refs.~\onlinecite{r_2015_cwe, r_2016_odg} and references therein.
Besides the usual bulk metals and diluted and concentrated alloys
containing heavy elements, 
large spin Hall currents can be induced by Rashba interface effects
\cite{r_2016_wwl, r_2017_bt, r_2018_azs, r_2019_lsx, r_2019_wgy}.
Further manifestation of spin-orbit effects in nonmagnetic solids
is a possible spin polarization of longitudinal currents due to
a special point group symmetry of the crystal structure
\cite{r_2015_wsc}.

Spin currents in systems with spontaneous magnetic moments
represent another wide area of the present intense research.
This field includes spin-transfer torques in layered systems
\cite{r_2008_rs}, spin polarization of longitudinal
\cite{r_2010_lke_b} and transversal \cite{r_2014_zck}
conductivities in bulk ferromagnetic alloys, as well as the SHE
in collinear \cite{r_2010_fbm} and noncollinear \cite{r_2017_zsy} 
antiferromagnets.
The latter systems have attracted special interest since 
their noncollinear spin structures can induce a sizable SHE
even without spin-orbit interaction \cite{r_2018_zzs}.

Reliable quantum-mechanical description of phenomena related to the
spin currents faces a basic problem due to the fact that the electron
spin in systems with spin-orbit interaction is not a conserving
quantity.
This hinders an exact definition of the spin-current operator;
fundamental approaches to this problem \cite{r_2006_wxs, r_2006_szx}
lead to expressions that cannot be employed directly in current
\emph{ab initio} techniques of electron theory of solids.
In this situation, existing practical solutions thus express the
spin-current operator typically as a symmetrized product of the spin
operator and the operator of charge current
\cite{r_2008_gmc, r_2010_lke_b, r_2016_wwl}.
It should be noted that the latter approaches are suitable even for
studies of disordered systems, in which the necessary configuration
averaging is performed either by a real-space supercell technique
\cite{r_2016_wwl} or by using the CPA
\cite{r_2011_lgk, r_2010_lke_b}.

The main aim of this work is a formulation of an alternative
first-principles approach to the spin currents in random alloys
which employs the idea of an intersite electron transport
\cite{r_2002_tkd}.
In this scheme, the intraatomic electron motion is systematically
neglected which leads to effective operators of charge current
that are spin-independent and nonrandom (independent of a
particular random configuration of the alloy).
The corresponding effective spin-current operators are nonrandom
as well, which allows us to define easily the spin-dependent
conductivity tensor and to perform its configuration averaging
within the single-site CPA \cite{r_1967_ps, r_1969_bv} in analogy
with the technique developed recently for the standard (charge)
conductivity tensor in relativistic theory
\cite{r_2012_tkd, r_2014_tkd}.

The paper is organized as follows.
The theoretical formalism is presented in Section~\ref{s_meth},
including the definition of the spin-dependent conductivity tensor
(Section~\ref{ss_sdctbf}), the CPA averaging of its
Fermi sea term (Section~\ref{ss_avsea}), a summary of the
formal properties of the derived theory (Section~\ref{ss_tpsdct}),
and details of numerical implementation (Section~\ref{ss_ind}).
Technical theoretical details are presented in the
Appendix~\ref{app_tisdct}.
The obtained results and their discussion, focused on the SHE in
nonmagnetic random alloys, are collected in Section~\ref{s_redi}.
First, a simple tight-binding model of a random binary alloy is
analyzed in the dilute limit in Section~\ref{ss_tbspm}.
Second, transport properties of selected Pt-based disordered fcc
alloys (Pt-Au, Pt-Re, Pt-Ta) are addressed in
Section~\ref{ss_ptball}.
Conclusions of the work are summarized in Section~\ref{s_conc}.

\section{Method\label{s_meth}}

\subsection{Spin-dependent conductivity tensor from 
            the Bastin formula\label{ss_sdctbf}} 

The starting point of our formalism is the Kubo linear response
theory \cite{r_1957_rk} and the formula of Bastin et al.\
\cite{r_1971_blb} for the full charge conductivity tensor
$\sigma_{\mu\nu}$ adapted in the relativistic tight-binding linear
muffin-tin orbital (TB-LMTO) method \cite{r_2014_tkd}
\begin{eqnarray}
\sigma_{\mu\nu} & = & 
-2\sigma_0 \int \mathrm{d}E f(E) \mathrm{Tr}
\left\langle v_\mu g'_+(E) v_\nu [ g_+(E) - g_-(E) ] \right.
\nonumber\\
 & & \left. \qquad \qquad \qquad \quad \
{} - v_\mu [ g_+(E) - g_-(E) ] v_\nu  g'_-(E) \right\rangle .
\label{eq_bastina}
\end{eqnarray}
In this relation, the quantity $\sigma_0 = e^2 \hbar/(4\pi \Omega)$,
where $\Omega$ is the volume of a big finite crystal with periodic
boundary conditions, $E$ is a real energy variable, $f(E)$ denotes
the Fermi-Dirac function, $g_\pm (E) = g(E \pm i 0)$ denote
side limits of the auxiliary Green's function $g(z)$ defined for
complex energies $z$, the prime at $g_\pm (E)$ denotes
energy derivative, the quantities $v_\mu$ ($\mu = x , y , z$)
are the effective velocity (current) operators, and the brackets
$\langle \dots \rangle$ refer to the configuration averaging for
random alloys.
The auxiliary Green's function is given as $g(z) = 
[ P(z) - S ]^{-1}$, where $P(z)$ represents a site-diagonal matrix
of the potential functions and $S$ denotes the structure constant
matrix.

The expression (\ref{eq_bastina}) reflects the intersite electron
transport which takes place inside the interstitial region among
the individual Wigner-Seitz cells (replaced by space-filling spheres
in the atomic-sphere approximation).
The operators $v_\mu$ and $g_\pm (E)$ are represented by matrices
in the composed index $\mathbf{R}L$, where $\mathbf{R}$ labels the
lattice sites and $L$ denotes the orbital index $L = (\ell m s)$
containing the orbital ($\ell$), magnetic ($m$), and spin ($s$)
quantum numbers ($s = \, \uparrow , \downarrow$).
The trace ($\mathrm{Tr}$) in Eq.~(\ref{eq_bastina}) refers to all
$\mathbf{R}L$-orbitals of the system.
Note that the orbital index $L$ corresponds to a nonrelativistic
theory despite that the fully relativistic solutions (including possible
spin polarization) of the single-site problem are used inside the
Wigner-Seitz cells (atomic spheres); this fact is due to the
nonrelativistic form of the LMTO orbitals in the interstitial
region (which reduce to spherical waves in a constant potential
with zero kinetic energy \cite{r_1975_oka, r_1984_aj}). 
The effective current operators $v_\mu$ are spin-independent and
nonrandom which follows from their definition \cite{r_2002_tkd,
r_2014_tkd} and from properties of the structure constant matrix
$S$.

The nonrelativistic character of the intersite electron transport
and the above properties of the effective current operators $v_\mu$
allow one to introduce naturally the effective spin-current
operators as $\sigma^\lambda v_\mu$, where the quantities 
$\sigma^\lambda$ ($\lambda = x , y , z$) equal the Pauli spin
matrices extended trivially to matrices in the composed
$\mathbf{R}L$ index, 
$(\sigma^\lambda)^{L' L}_{\mathbf{R}' \mathbf{R}} =
\delta_{\mathbf{R}' \mathbf{R}} \delta_{\ell' \ell} \delta_{m' m}
\sigma^\lambda_{s' s}$ where $L' = (\ell' m' s')$.
This definition represents an analogy of spin-current operators
employed in other studies
\cite{r_2008_gmc, r_2010_lke_b, r_2016_wwl}.
The spin-dependent conductivity tensor $\sigma^\lambda_{\mu\nu}$
corresponding to the original charge conductivity tensor
(\ref{eq_bastina}) is then defined by
\begin{eqnarray}
\sigma^\lambda_{\mu\nu} & = & 
-2\sigma_0 \int \mathrm{d}E f(E) \mathrm{Tr}
\left\langle ( \sigma^\lambda v_\mu ) g'_+(E) v_\nu 
[ g_+(E) - g_-(E) ] \right.
\nonumber\\
 & & \left. \qquad \qquad \qquad \quad \
{} - ( \sigma^\lambda v_\mu )
[ g_+(E) - g_-(E) ] v_\nu  g'_-(E) \right\rangle ,
\label{eq_bastinh}
\end{eqnarray}
which describes the linear response of the spin current 
$\sigma^\lambda v_\mu$ to a spin-independent electrical field
in the direction of the $\nu$ axis.
Alternatively, one can consider the response coefficient
\begin{eqnarray}
\tilde{\sigma}_{\mu\nu} & = & 
-2\sigma_0 \int \mathrm{d}E f(E) \mathrm{Tr}
\left\langle \tilde{v}_\mu g'_+(E) v_\nu 
[ g_+(E) - g_-(E) ] \right.
\nonumber\\
 & & \left. \qquad \qquad \qquad \quad \
{} - \tilde{v}_\mu 
[ g_+(E) - g_-(E) ] v_\nu  g'_-(E) \right\rangle ,
\label{eq_bastink}
\end{eqnarray}
where $\tilde{v}_\mu = ({\bf n} \cdot \bm{\sigma}) v_\mu$ denotes
the spin-polarized effective velocity (current) with the
spin-polarization axis along a global nonrandom unit vector
${\bf n}$. 
Note that the spin-polarized velocities $\tilde{v}_\mu$ are
nonrandom operators, which simplifies the configuration
averaging in Eq.~(\ref{eq_bastink}).

In full analogy to $\sigma_{\mu\nu}$, the spin-dependent
conductivity tensor $\tilde{\sigma}_{\mu\nu}$ can be decomposed
into a Fermi surface term and a Fermi sea term as
\cite{r_2014_tkd, r_2001_cb} 
\begin{equation}
\tilde{\sigma}_{\mu\nu} = \tilde{\sigma}^{(1)}_{\mu\nu} 
 + \tilde{\sigma}^{(2)}_{\mu\nu} .
\label{eq_sigma}
\end{equation}
For systems at zero temperature, the Fermi surface term 
$\tilde{\sigma}^{(1)}_{\mu\nu}$ can be written as
\begin{eqnarray}
\tilde{\sigma}^{(1)}_{\mu\nu} & = & \sigma_0 \mathrm{Tr} 
\left\langle \tilde{v}_\mu [ g_+(E_\mathrm{F}) - g_-(E_\mathrm{F}) ] 
v_\nu  g_-(E_\mathrm{F}) \right.
\nonumber\\
 & & \left. \quad \
{} - \tilde{v}_\mu g_+(E_\mathrm{F}) v_\nu [ g_+(E_\mathrm{F}) 
 - g_-(E_\mathrm{F}) ] \right\rangle ,
\label{eq_surf}
\end{eqnarray}
where $E_\mathrm{F}$ denotes the Fermi energy.
The Fermi sea term $\tilde{\sigma}^{(2)}_{\mu\nu}$ can be
reformulated as a complex contour integral
\begin{equation}
\tilde{\sigma}^{(2)}_{\mu\nu} = \sigma_0 \int_C \mathrm{d}z  
\mathrm{Tr} \langle \tilde{v}_\mu g'(z) v_\nu g(z) 
- \tilde{v}_\mu g(z) v_\nu g'(z) \rangle ,
\label{eq_seaj}
\end{equation}
where the integration path $C$ starts and ends at $E_\mathrm{F}$,
it is oriented counterclockwise and it encompasses the whole occupied
part of the alloy valence spectrum.

The configuration average in the CPA of the Fermi surface term
yields its coherent part (coh) and the incoherent part (vertex
corrections -- VC), $\tilde{\sigma}^{(1)}_{\mu\nu} =
\tilde{\sigma}^{(1)}_{\mu\nu,\mathrm{coh}} +
\tilde{\sigma}^{(1)}_{\mu\nu,\mathrm{VC}}$, where
\begin{eqnarray}
\tilde{\sigma}^{(1)}_{\mu\nu,\mathrm{coh}} & = & 
\sigma_0 \mathrm{Tr} \left\{ \tilde{v}_\mu
 [ \bar{g}_+(E_\mathrm{F}) - \bar{g}_-(E_\mathrm{F}) ] 
v_\nu  \bar{g}_-(E_\mathrm{F}) \right.
\nonumber\\
 & & \left. \quad \
{} - \tilde{v}_\mu \bar{g}_+(E_\mathrm{F}) v_\nu 
[ \bar{g}_+(E_\mathrm{F}) - \bar{g}_-(E_\mathrm{F}) ] \right\} ,
\label{eq_surfcoh}
\end{eqnarray}
while the vertex corrections
$\tilde{\sigma}^{(1)}_{\mu\nu,\mathrm{VC}}$ are evaluated according
to the original CPA theory \cite{r_1969_bv} adapted to the TB-LMTO
formalism \cite{r_2006_ctk}.
The symbols $\bar{g}_\pm (E)$ in Eq.~(\ref{eq_surfcoh}) and
$\bar{g} (z)$ in the following text denote the configuration averages
of $g_\pm (E)$ and $g(z)$, respectively.
These quantities are given by $\bar{g} (z) = 
[ \mathcal{P}(z) - S ]^{-1}$, where $\mathcal{P}(z)$ is a
site-diagonal matrix of the coherent potential functions.
The treatment of the Fermi sea term (\ref{eq_seaj}) is done similarly
to the case of the charge conductivity tensor \cite{r_2014_tkd}; 
the details are given in Section~\ref{ss_avsea}.

\subsection{Configuration averaging of the Fermi sea
            term\label{ss_avsea}} 

In analogy with our recent study \cite{r_2014_tkd}, we find that
the CPA average of the Fermi sea term (\ref{eq_seaj}) can be
simplified owing to the exact vanishing of the on-site blocks of the
matrix product $\bar{g}(z) v_\nu \bar{g}(z)$:  
\begin{equation}
[ \bar{g}(z) v_\nu \bar{g}(z) ]^{LL'}_{\mathbf{R} \mathbf{R}} = 0 ,
\label{eq_gvgon}
\end{equation}
which is valid for the same energy arguments of both Green's 
functions.
This rule is a consequence of a simple form of the underlying
coordinate operators and of the single-site nature of the coherent
potential functions $\mathcal{P}(z)$, see
Ref.~\onlinecite{r_2014_tkd} for details.
Following the procedure outlined previously \cite{r_2014_tkd}, one
can derive the resulting averages in Eq.~(\ref{eq_seaj}) as
\begin{equation}
\mathrm{Tr} \langle \tilde{v}_\mu g' v_\nu g \rangle =
\mathrm{Tr} \{ \tilde{v}_\mu \bar{g}' v_\nu \bar{g} \} 
 + \sum_{\mathbf{R}_1 \Lambda_1} \sum_{\mathbf{R}_2 \Lambda_2} 
\left[ \bar{g} \tilde{v}_\mu \bar{g} 
\right]^{L'_1 L_1}_{\mathbf{R}_1 \mathbf{R}_1}
\left[ \Delta^{-1} 
\right]^{\Lambda_1 \Lambda_2}_{\mathbf{R}_1 \mathbf{R}_2}
\left[ \bar{g}' v_\nu \bar{g} 
\right]^{L_2 L'_2}_{\mathbf{R}_2 \mathbf{R}_2} 
\label{eq_vgpvgf}
\end{equation}
and
\begin{equation}
\mathrm{Tr} \langle \tilde{v}_\mu g v_\nu g' \rangle =
\mathrm{Tr} \{ \tilde{v}_\mu \bar{g} v_\nu \bar{g}' \} 
 + \sum_{\mathbf{R}_1 \Lambda_1} \sum_{\mathbf{R}_2 \Lambda_2} 
\left[ \bar{g} \tilde{v}_\mu \bar{g} 
\right]^{L'_1 L_1}_{\mathbf{R}_1 \mathbf{R}_1}
\left[ \Delta^{-1} 
\right]^{\Lambda_1 \Lambda_2}_{\mathbf{R}_1 \mathbf{R}_2}
\left[ \bar{g} v_\nu \bar{g}' 
\right]^{L_2 L'_2}_{\mathbf{R}_2 \mathbf{R}_2} ,
\label{eq_vgvgpf}
\end{equation}
where all energy arguments of the matrices $g(z)$, $g'(z)$ and
$\Delta(z,z)$ have been omitted for brevity.
The symbols $\Lambda_1$ and $\Lambda_2$ abbreviate the composed
orbital indices $\Lambda_1 = ( L_1 , L'_1 )$ and
$\Lambda_2 = ( L_2 , L'_2 )$, respectively, and the matrix 
$\Delta^{\Lambda_1 \Lambda_2}_{\mathbf{R}_1 \mathbf{R}_2}
( z_1 , z_2 )$ was defined in the Appendix of
Ref.~\onlinecite{r_2006_ctk}.
The first terms in (\ref{eq_vgpvgf}) and (\ref{eq_vgvgpf}) define
the coherent contributions whereas the second terms are the
corresponding vertex corrections; note that these vertex parts
differ mutually only by their signs as a direct consequence of the
rule~(\ref{eq_gvgon}).

The resulting coherent (coh) and vertex (VC) contributions to the
Fermi sea term $\tilde{\sigma}^{(2)}_{\mu\nu} =
\tilde{\sigma}^{(2)}_{\mu\nu,\mathrm{coh}} +
\tilde{\sigma}^{(2)}_{\mu\nu,\mathrm{VC}}$ are given from the
respective terms in (\ref{eq_vgpvgf}) and (\ref{eq_vgvgpf}):
\begin{equation}
\tilde{\sigma}^{(2)}_{\mu\nu,\mathrm{coh}} = 
\sigma_0 \int_C \mathrm{d}z  \mathrm{Tr} \left\{ \tilde{v}_\mu
\bar{g}'(z) v_\nu \bar{g}(z) - \tilde{v}_\mu \bar{g}(z) 
v_\nu \bar{g}'(z) \right\} ,
\label{eq_seacoh}
\end{equation}
and
\begin{equation}
\tilde{\sigma}^{(2)}_{\mu\nu,\mathrm{VC}} = 
2 \sigma_0 \int_C \mathrm{d}z 
\sum_{\mathbf{R}_1 \Lambda_1} \sum_{\mathbf{R}_2 \Lambda_2} 
\left[ \bar{g} \tilde{v}_\mu \bar{g} 
\right]^{L'_1 L_1}_{\mathbf{R}_1 \mathbf{R}_1}
\left[ \Delta^{-1} 
\right]^{\Lambda_1 \Lambda_2}_{\mathbf{R}_1 \mathbf{R}_2}
\left[ \bar{g}' v_\nu \bar{g} 
\right]^{L_2 L'_2}_{\mathbf{R}_2 \mathbf{R}_2} ,
\label{eq_seavc}
\end{equation}
where the energy argument $z$ in $\bar{g}(z)$, $\bar{g}'(z)$ and
$\Delta(z,z)$ has been suppressed.
The appearance of the incoherent part of the Fermi sea
term~(\ref{eq_seavc}) represents the main difference between the
spin-dependent and standard conductivity tensors in the TB-LMTO-CPA
formalism, since the Fermi sea term in $\sigma_{\mu\nu}$
is purely coherent \cite{r_2014_tkd}.

The energy derivative of the auxiliary Green's function, encountered
in (\ref{eq_seacoh}) and (\ref{eq_seavc}), is obtained from the rule
$\bar{g}'(z) = - \bar{g}(z) \mathcal{P}'(z) \bar{g}(z)$, which
follows from the energy independent structure constants $S$.
As discussed in detail in Ref.~\onlinecite{r_2014_tkd}, the
formulation of the energy derivative $\mathcal{P}'(z)$ leads to
CPA-vertex corrections involving an inversion of the same kernel
$\Delta(z,z)$ as in $\tilde{\sigma}^{(2)}_{\mu\nu,\mathrm{VC}}$
(\ref{eq_seavc}).
This simplifies the numerical evaluation.

Finally, let us note that the derived formulas for the
spin-dependent conductivity tensor (in this and previous section)
are valid not only for spin currents as observables, but they
represent a general TB-LMTO-CPA result for the response of any
quantity to an applied electric field. 
The developed formalism can thus be directly used, e.g., to
calculate the torkance tensor relevant for spin-orbit torques
induced by electric fields \cite{r_2009_mz, r_2014_fbm, r_2016_wcs}.

\subsection{Transformation properties of the spin-dependent
            conductivity tensor\label{ss_tpsdct}} 

Since the enhanced numerical efficiency of the TB-LMTO method
as compared to the original (canonical) LMTO technique is due to
the screening of the structure constants, which depends on the
chosen LMTO representation $\alpha$ \cite{r_1984_aj, r_1986_apj},
the invariance of all physical quantities with respect to $\alpha$
is a necessary condition for any proper theoretical formalism.
In the context of relativistic transport properties, this check
has been done in detail for the standard conductivity tensor 
$\sigma_{\mu\nu}$ (\ref{eq_bastina}) in Ref.~\onlinecite{r_2014_tkd}
and for the Gilbert damping tensor in Ref.~\onlinecite{r_2015_tkd}. 

In the case of the spin-dependent conductivity tensor
$\tilde{\sigma}_{\mu\nu}$ (\ref{eq_bastink}) and of its Fermi
surface and Fermi sea terms, the detailed study is outlined in
the Appendix \ref{app_tisdct}.
Here we merely list the quantities invariant with respect to the
choice of LMTO representation:
(i) the total tensor $\tilde{\sigma}_{\mu\nu}$, 
(ii) the sum of the coherent contributions to the Fermi surface and
Fermi sea terms, i.e., 
$\tilde{\sigma}^{(1)}_{\mu\nu,\mathrm{coh}} 
+\tilde{\sigma}^{(2)}_{\mu\nu,\mathrm{coh}}$,
(iii) the vertex corrections to the Fermi surface term,
$\tilde{\sigma}^{(1)}_{\mu\nu,\mathrm{VC}}$, and
(iv) the vertex corrections to the Fermi sea term,
$\tilde{\sigma}^{(2)}_{\mu\nu,\mathrm{VC}}$.

In analogy with the theory of the charge conductivity tensor
\cite{r_2014_tkd}, the above invariant contributions to the total
spin-dependent conductivity tensor $\tilde{\sigma}_{\mu\nu}$ can be
used for the definition of its intrinsic part, given by the sum
of the coherent terms $\tilde{\sigma}^{(1)}_{\mu\nu,\mathrm{coh}}
+\tilde{\sigma}^{(2)}_{\mu\nu,\mathrm{coh}}$, and of its extrinsic
part, equal to the sum of the incoherent terms (vertex corrections)
$\tilde{\sigma}^{(1)}_{\mu\nu,\mathrm{VC}} +
\tilde{\sigma}^{(2)}_{\mu\nu,\mathrm{VC}}$.
Since the tensor $\tilde{\sigma}_{\mu\nu}$ contains the SHC, this
separation is naturally extended also to the intrinsic and extrinsic
SHC.
The extrinsic SHC is dominated by the Fermi surface contribution,
which includes a skew-scattering term (diverging in the dilute
limit) and a side-jump term (approaching a finite value in the
same limit),
whereas the Fermi sea contribution is practically negligible
both in diluted and concentrated alloys, see Section~\ref{s_redi}.
This classification is essentially identical to that adopted in
previous \emph{ab initio} theories of the SHE \cite{r_2010_gfz,
r_2011_lgk, r_2015_cwe}.

A more detailed separation of the terms due to individual mechanisms
of the SHC (skew scattering, side jump) has recently been considered
by other authors.
For diluted alloys, a practical procedure has been suggested which
rests on asymptotic concentration dependences (in the dilute limit)
of the individual SHC contributions and of the longitudinal
conductivity \cite{r_2015_cfh}.
In the case of concentrated alloys, an alternative scheme has been
worked out by the authors of Ref.~\onlinecite{r_2016_hsk} for the
anomalous Hall effect with simplified spin-orbit coupling
(of $\xi \bm{L} \cdot \bm{S}$ form).
This approach employs an energy-dependent nonrandom TB-LMTO-CPA
Hamiltonian \cite{r_1990_kd} whose eigenvectors allow one to define
interband and intraband matrix elements of the velocity operators
needed for the separation.
A generalization of this procedure to the SHC in the fully
relativistic theory remains yet to be done.

\subsection{Implementation and numerical details\label{ss_ind}} 

The numerical implementation of the developed formalism and the
performed calculations follow closely our recent works focused on
the Fermi surface \cite{r_2012_tkd} and Fermi sea \cite{r_2014_tkd}
terms of the charge conductivity tensor.
We have employed the $spd$ basis of the selfconsistent relativistic
TB-LMTO-CPA method \cite{r_1997_tdk},
added a small imaginary part of $\pm 10^{-5}$ Ry to the Fermi energy
$E_\mathrm{F}$ in evaluation of $\tilde{\sigma}^{(1)}_{\mu\nu}$, and
used 20 -- 40 complex nodes for integrations along the complex
contour $C$ in evaluation of $\tilde{\sigma}^{(2)}_{\mu\nu}$.
The Brillouin zone integrals were performed with sufficient numbers
of $\mathbf{k}$ points;
for the complex energy arguments closest to the real Fermi energy,
total numbers of $\sim 10^8$ sampling points were used.

\section{Results and discussion\label{s_redi}}

The first results of the developed theory, discussed in this paper,
refer to nonmagnetic random binary alloys on fcc lattices.
As a consequence of the full cubic symmetry and time-inversion
symmetry \cite{r_2015_wsc}, the only independent nonzero element
of the tensor $\sigma^\lambda_{\mu\nu}$ (\ref{eq_bastinh}) is
$\sigma^z_{xy}$ which is equivalent to $\tilde{\sigma}_{xy}$
(\ref{eq_bastink}) with the spin-polarization vector $\mathbf{n}$
along the $z$ axis. 
This element is identified with the SHC in the following.

\subsection{Random alloy in a tight-binding
            $sp$ model\label{ss_tbspm}}

\begin{figure}[ht]
\begin{center}
\includegraphics[width=0.50\textwidth]{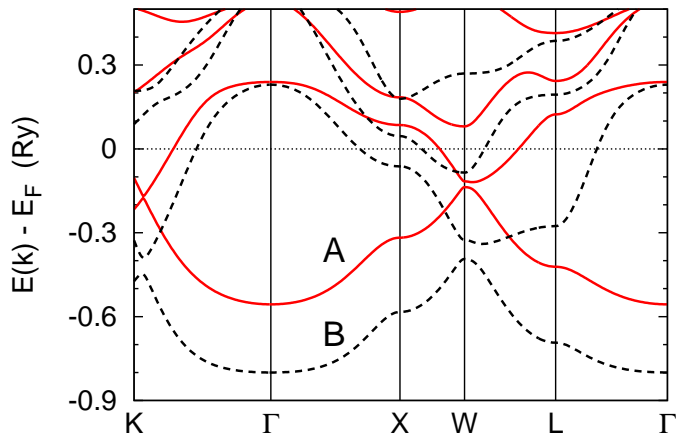}
\end{center}
\caption{
Band structures of pure metals A (full lines) and B (dashed
lines) in a tight-binding $sp$ model on the fcc lattice.
The horizontal dotted line denotes the position of the
Fermi energy.
\label{f_tbbs}}
\end{figure}

In order to investigate the behavior of the three contributions
to the SHC (Section~\ref{ss_tpsdct}) in a dilute limit, we have
studied a simple tight-binding model of a random binary alloy
A$_{1-c}$B$_c$ on an fcc lattice.
The model assumes $sp$ orbitals on each lattice site with
a site-diagonal disorder present in the LMTO potential parameters
of both alloy constituents, see Table~\ref{tablep};
the band structures of ideal fcc metals A and B are displayed in
Fig.~\ref{f_tbbs}.
The species A is lighter than the species B, which is reflected
by higher eigenvalues of A as compared to those of B.
Note that only one band (including its double degeneracy)
intersects the Fermi energy for metal A, whereas two bands cross
the $E_\mathrm{F}$ for metal B.
Simultaneously, the strength of spin-orbit coupling of A is
smaller than that of B which is documented, e.g., by a smaller
splitting of the two lowest bands of A at the point W as compared
to the corresponding splitting of B.

\begin{table}[htb]
\caption{\label{tablep}
The LMTO potential parameters $C_\kappa$, $\Delta^{1/2}_\kappa$,
and $\gamma_\kappa$ for three values of the relativistic index
$\kappa$ and for both species of the model A-B alloy
(values in parentheses refer to the impurity atom B). 
The Wigner-Seitz radius was set to $3.0\, a_0$, where $a_0$ denotes
the Bohr radius, and the Fermi energy was set to zero.}
\begin{ruledtabular}
\begin{tabular}{cccc}
$\kappa$  & $C_\kappa$ (Ry) & $\Delta^{1/2}_\kappa$ (Ry$^{1/2}$) 
          & $\gamma_\kappa$ \\
\colrule
 $-1$  & $-0.25\, (-0.60)$ & $0.35\, (0.30)$ & $0.40\, (0.45)$ \\
 $-2$  & $-0.05\, (-0.25)$ & $0.25\, (0.27)$ & $0.06\, (0.08)$ \\
 $+1$  & $-0.10\, (-0.35)$ & $0.20\, (0.25)$ & $0.05\, (0.06)$ \\
\end{tabular}
\end{ruledtabular}
\end{table}

\begin{figure}[ht]
\begin{center}
\includegraphics[width=0.45\textwidth]{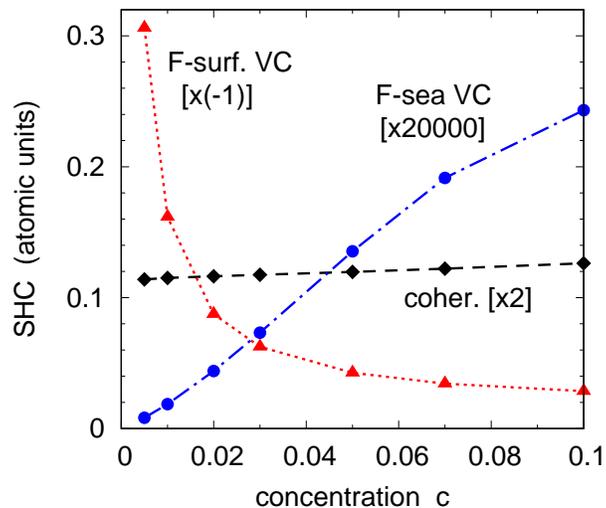}
\end{center}
\caption{
Three contributions to the spin Hall conductivity (SHC) for
a tight-binding $sp$ model of a random fcc alloy A$_{1-c}$B$_c$
as functions of the concentration $c$:
the coherent term (diamonds), the vertex corrections to the
Fermi surface term (triangles), and the vertex corrections
to the Fermi sea term (circles).
These contributions were rescaled by a factor of 2, $-1$,
and 20000, respectively. 
\label{f_tbshc}}
\end{figure}

In this simple model of a random A$_{1-c}$B$_c$ alloy, the fcc
lattice parameter, the species-resolved LMTO potential parameters,
and the alloy Fermi energy were kept fixed (independent on the
concentration $c$).
The resulting SHC contributions in the dilute alloy for $c \to 0$
are shown in Fig.~\ref{f_tbshc}; the SHC values are given in
atomic units which is sufficient for the present purpose.
One can see that the three contributions exhibit three different
concentration trends for the vanishing impurity content:
the coherent term ($\tilde{\sigma}^{(1)}_{xy,\mathrm{coh}}
+\tilde{\sigma}^{(2)}_{xy,\mathrm{coh}}$) exhibits a finite
nonzero limit, the Fermi surface vertex term 
($\tilde{\sigma}^{(1)}_{xy,\mathrm{VC}}$) diverges with an
inverse proportionality to $c$, and the Fermi sea vertex term 
($\tilde{\sigma}^{(2)}_{xy,\mathrm{VC}}$) vanishes roughly
linearly with $c$.

The divergence of the Fermi surface vertex term has been obtained
and discussed by a number of authors \cite{r_2015_svw, r_2010_gfz,
r_2011_lgk}; this behavior of the extrinsic SHC has been ascribed
to skew scattering.
The weakly concentration dependent coherent term has also been found
earlier \cite{r_2011_lgk, r_2015_cwe}.
This trend justifies an identification of the coherent term with
the intrinsic SHC of random alloys.
Let us note however that the limit of the coherent SHC term for
$c \to 0$ does not necessarily coincide with the SHC of the pure
host metal A evaluated by using the Berry curvature approach from
its band structure.
This fact, ascribed to the coherent part of the side-jump
contribution \cite{r_2015_svw, r_2015_cfh},
can be explained by a sensitivity of the alloy selfenergy to the
impurity B potential which affects the limiting value of the
coherent SHC for $c \to 0$.
A complete analysis of this point goes beyond the scope of this work.

The vertex corrections to the Fermi sea term have not been
explicitly studied by other authors in the dilute limit.
Since the same kernel $\Delta(z,z)$ is inverted in evaluation of
Eq.~(\ref{eq_seavc}) and in obtaining the energy derivative of the
coherent potential function $\mathcal{P}'(z)$, the revealed
proportionality $\tilde{\sigma}^{(2)}_{xy,\mathrm{VC}} \sim c$ can
be understood as a counterpart of the limiting behavior
$\mathcal{P}'(z) \to P'_\mathrm{A}(z)$ for $c \to 0$, where
$P_\mathrm{A}(z)$ denotes the potential function of the host metal A.
Note however that the magnitude of the incoherent Fermi sea term is
very small as compared to the other two terms (Fig.~\ref{f_tbshc});
since similarly tiny magnitudes were found for realistic alloy models
even in concentrated regimes (Section~\ref{ss_ptball}), a more
detailed discussion of this term seems to be of little importance.

\subsection{Random fcc Pt-based alloys\label{ss_ptball}}

In this section, we address random fcc alloys of Pt with other heavy
metals Au, Re, and Ta.
For all systems, the average Wigner-Seitz (atomic sphere) radius of
the alloy was set according to the Vegard's law and the experimental
values of the atomic sphere radii of the pure elements in their
equilibrium structures.
Local lattice relaxations were ignored which is acceptable because 
of similar sizes of all four elements.

\begin{figure}[ht]
\begin{center}
\includegraphics[width=0.80\textwidth]{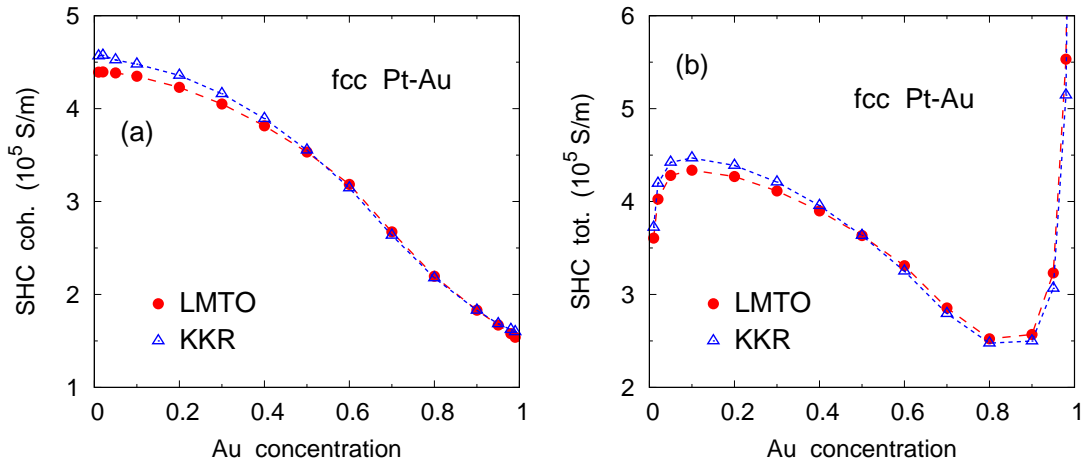}
\end{center}
\caption{
Comparison of the spin Hall conductivities (SHC) calculated
in the TB-LMTO (solid circles) and KKR (open triangles) techniques
for random fcc Pt-Au alloys:
(a) the coherent part of the SHC and (b) the total SHC, both as
functions of Au concentration.
The results of the KKR method were taken from
Ref.~\onlinecite{r_2016_odg}.
\label{f_ptaukkr}}
\end{figure}

The total SHC and its intrinsic part for Pt-Au alloys are
shown in Fig.~\ref{f_ptaukkr}; the TB-LMTO results are compared with
those of the relativistic Korring-Kohn-Rostoker (KKR) method
\cite{r_2016_odg}.
One can see a very good agreement between both methods taking into
account the different spin-current operators used: random,
site-diagonal operators enter the KKR technique \cite{r_2010_lke_b},
whereas nonrandom, non-site-diagonal effective operators are
employed in the present work (Section~\ref{ss_sdctbf}).
The intrinsic SHC decreases monotonically with increasing Au 
concentration (Fig.~\ref{f_ptaukkr}a); the extrinsic contribution
modifies this trend significantly only near the limits of pure Pt
and pure Au (Fig.~\ref{f_ptaukkr}b), where the divergent behavior
dominates.
For all concentrations, the vertex corrections to the Fermi sea term
are about four orders of magnitude smaller than the SHC values in
Fig.~\ref{f_ptaukkr} and can thus be safely ignored.
This feature has also been found in the KKR results \cite{r_2015_kce}.

The agreement between the results obtained by the TB-LMTO and the
KKR techniques calls for a deeper theoretical explanation.
Disregarding various technical details of both approaches, one
finds that the most profound difference rests in the use of the
usual, continuous coordinates in the KKR method and of the modified,
steplike coordinates (constant inside each Wigner-Seitz cell) in
the TB-LMTO method \cite{r_2002_tkd}.
One can then prove that the two different current (velocity)
operators lead to the same values of all elements of the standard
conductivity tensor $\sigma_{\mu\nu}$.
The proof employs relations between isothermic and adiabatic
linear-response coefficients as well as properties of the current
operator for systems in thermodynamic equilibrium, see
Ref.~\onlinecite{r_sumact} for more details.
We are unable to provide a similar proof for the spin-dependent
conductivity tensor $\tilde{\sigma}_{\mu\nu}$;
let us note that an equivalence of different torque operators for
the Gilbert damping tensor in random ferromagnets has been shown
elsewhere \cite{r_2015_tkd, r_2017_dtk}.
Comparison of both approaches from a practical (numerical) point
of view reveals essentially the same efficiency and accuracy in
evaluation of the Fermi surface term of the conductivity tensors.
The Fermi sea term, which involves energy derivatives of the Green's
functions, is treated by means of numerical derivatives in the KKR
method \cite{r_2015_kce}, whereas analytical relations (owing to the
parametrized potential functions and energy-independent structure
constants) are used for these derivatives in the TB-LMTO method
which seems advantageous for the computations.

\begin{figure}[ht]
\begin{center}
\includegraphics[width=0.80\textwidth]{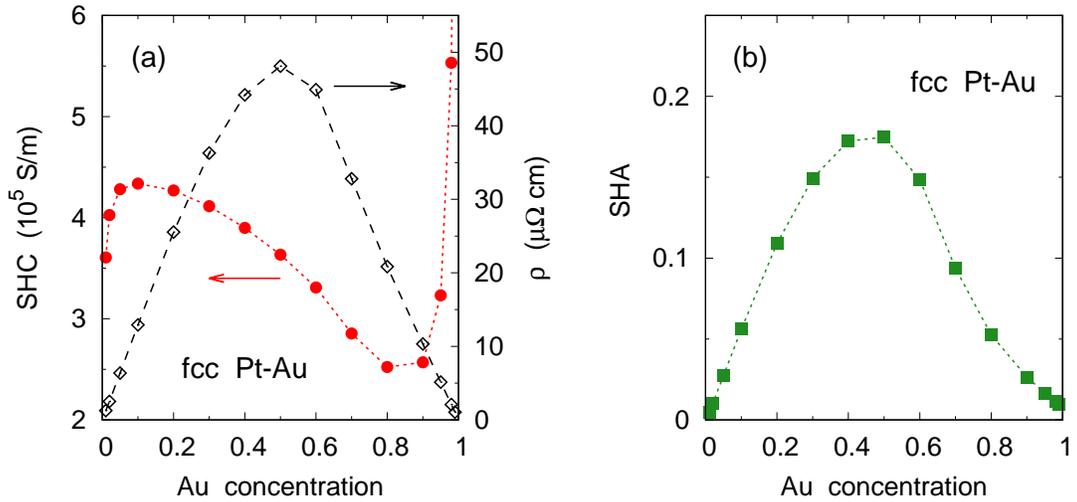}
\end{center}
\caption{
The calculated spin Hall conductivities (SHC), residual
resistivities ($\rho$), and spin Hall angles (SHA)
in random fcc Pt-Au alloys as functions of Au concentration:
(a) SHC (solid circles, left scale) and $\rho$ (open diamonds,
right scale) and (b) SHA (solid squares).
\label{f_ptausha}}
\end{figure}

The calculated SHC values for Pt-Au alloys compare reasonably well
to the measured ones in the entire concentration interval
\cite{r_2016_odg}.
Figure~\ref{f_ptausha} displays the concentration trend of the SHC
and of other transport quantities: the longitudinal resistivity
$\rho$ and the SHA given by the product $\tilde{\sigma}_{xy} \rho$.
One can see a maximum of $\rho$ for the equiconcentration alloy
(Fig.~\ref{f_ptausha}a); a very similar concentration trend is
obtained for the SHA with a maximum value slightly below 0.2
(Fig.~\ref{f_ptausha}b) which agrees again with the KKR results
\cite{r_2016_odg}.
This value is comparable with the top SHA values obtained for other
alloys based on $5d$ transition metals \cite{r_2016_odg, r_2018_fwe}.
Note however that the equiconcentration Pt-Au system is not
thermodynamically stable at ambient temperatures according to its
equilibrium phase diagram \cite{r_1986_tbm} so that the measured
samples are stabilized only by kinetic barriers.
This fact calls for inspection of other alloy systems which might
exhibit sizable SHA values for substitutional solid solutions that
are equilibrium phases at low temperatures.

\begin{figure}[ht]
\begin{center}
\includegraphics[width=0.80\textwidth]{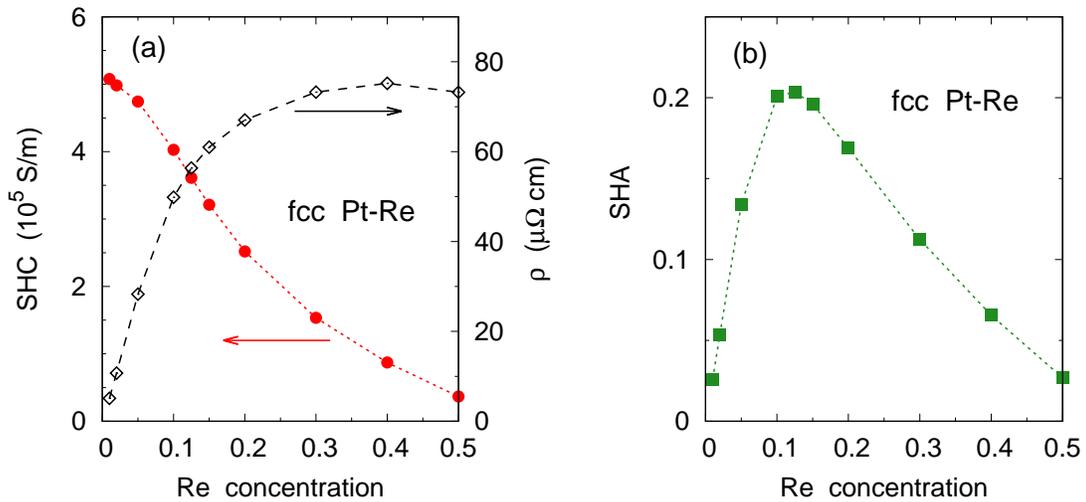}
\end{center}
\caption{
The same as Fig.~\ref{f_ptausha} but for random fcc Pt-Re alloys.
\label{f_ptresha}}
\end{figure}

In this work, we confined our interest to Pt-based alloys with
fcc structure.
The solubility limit of Re in fcc Pt is about 40 at.\% Re
\cite{r_1986_tbm}.
The calculated values of SHC, $\rho$, and SHA are shown in
Fig.~\ref{f_ptresha}.
One can see a decreasing trend of the SHC due to alloying by Re,
which is accompanied by a steep increase of the resistivity $\rho$
for small Re contents followed by a saturation of $\rho$ for higher
Re concentrations (Fig.~\ref{f_ptresha}a).
As a result of these trends, the SHA exhibits a maximum value of
about 0.2 for Re content around 12 at.\% Re (Fig.~\ref{f_ptresha}b).
This composition falls safely inside the solubility interval of this
alloy system and the predicted SHA value thus might deserve future
experimental verification.

\begin{figure}[ht]
\begin{center}
\includegraphics[width=0.80\textwidth]{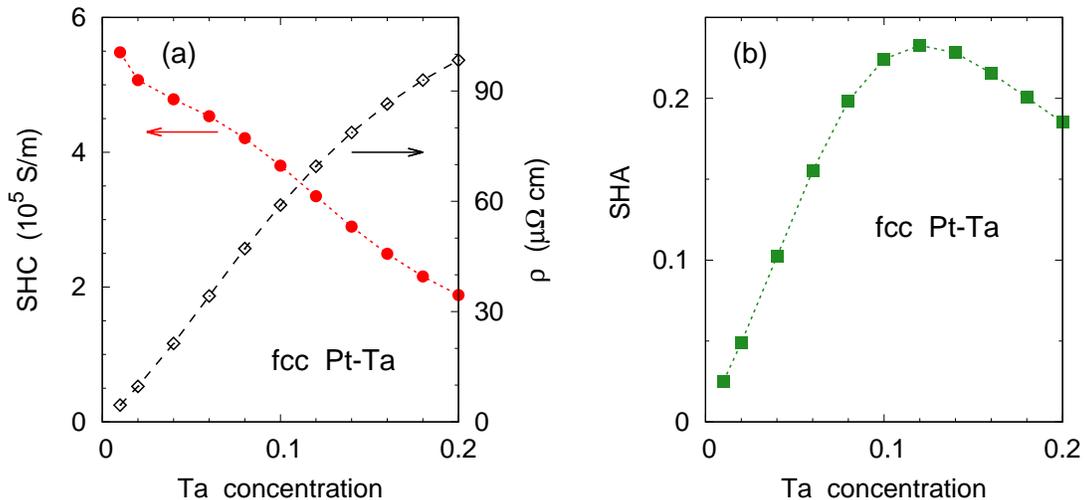}
\end{center}
\caption{
The same as Fig.~\ref{f_ptausha} but for random fcc Pt-Ta alloys.
\label{f_pttasha}}
\end{figure}

The phase diagram of the Pt-rich Pt-Ta system at ambient temperatures
is not exactly known at present; extrapolations from higher
temperatures indicate that the low-temperature solubility limit
is about 15 at.\% Ta \cite{r_1986_tbm}.
The relevant collection of calculated SHC, $\rho$, and SHA is
displayed in Fig.~\ref{f_pttasha}.
One can observe opposite trends of SHC and $\rho$ due to alloying
by Ta (Fig.~\ref{f_pttasha}a), in analogy with the Pt-Au
(Fig.~\ref{f_ptausha}a) and Pt-Re (Fig.~\ref{f_ptresha}a) systems.
The SHA exhibits a maximum again; the maximum SHA exceeds slightly
the value 0.2 which is obtained for the alloy with about 12 at.\% Ta
(Fig.~\ref{f_pttasha}b).
This composition should correspond to a thermodynamically stable
primary solid solution.

The obtained maximum SHA values in all three studied Pt-based alloys
result from a delicate competition between a reduction of SHC and an
increase of $\rho$ due to alloying.
Since these maximum values increase in sequence Pt-Au, Pt-Re, and
Pt-Ta, one can ascribe a more important role to variations of $\rho$
which increases in the same sequence, as can be seen, e.g., by
comparing the resistivities for alloys with 20 at.\% of impurities.
This importance of the longitudinal resistivity for large SHA values
is in line with recent findings for the SHE in amorphous Hf-W alloys
\cite{r_2018_fwe} as well as for SHE-based spin torques in
multilayers with Pt-Al and Pt-Hf alloys \cite{r_2016_nzr}.
Moreover, the resistivity variation is also responsible for the
temperature dependence of the SHA both in bulk Pt-Au alloys
\cite{r_2016_odg} and in Pt layers adjacent to ferromagnetic Ni-Fe
alloys \cite{r_2016_wwl}.

\section{Conclusions\label{s_conc}}

We have modified our recent theory of the relativistic electron
transport in random alloys within the TB-LMTO-CPA method
\cite{r_2014_tkd} for the spin-dependent conductivity tensor.
The derived formalism leads in general to three contributions to
this tensor, namely (i) a coherent part, (ii) an incoherent part
of the Fermi surface term, and (iii) an incoherent part of the
Fermi sea term, which are all invariant with respect to the chosen
TB-LMTO representation.
The coherent part can be identified with an intrinsic contribution
to the tensor, whereas the incoherent parts lead to an extrinsic
contribution.

The developed theory is in principle applicable to a wide spectrum
of phenomena involving spin-polarized currents induced by external
electric fields encountered in systems with or without spontaneous
magnetic moments.
The performed analysis and calculations related to the spin Hall
effect in nonmagnetic alloys revealed that the three contributions
to the spin Hall conductivity exhibit three different concentration
trends in the dilute limit and that the incoherent part of the Fermi
sea term is negligibly small in the entire concentration interval.
The obtained results for selected Pt-based binary alloys indicate
that sizable values of the spin Hall angles can be obtained even
for thermodynamically equilibrium primary solid solutions as an
alternative to often studied metastable crystalline and amorphous
alloys.

The approach worked out in this paper is not restricted only to the
spin currents as observables; the derived formulas for the
spin-dependent conductivity tensor represent essentially a complete
result (within the TB-LMTO-CPA method) for the static linear response
of any quantity to an external electric field.
For this reason, a possible extension of the present theory towards
a treatment of spin-orbit torques due to electric fields
\cite{r_2009_mz, r_2014_fbm, r_2016_wcs} seems (with the use of
nonrandom effective torque operators \cite{r_2015_tkd}) quite
promising.

\begin{acknowledgments}
The authors acknowledge financial support from the Czech Science
Foundation (Grant No.\ 18-07172S).
\end{acknowledgments}


%\appendix*
\appendix

\section{Transformation invariance of the spin-dependent
         conductivity tensor\label{app_tisdct}}

\subsection{One-particle quantities \label{app_opq}}

The study of the invariance of physical quantities with respect to
the choice of the LMTO representation is based on relations for
the coherent potential functions and the structure constants in
two different representations, denoted by superscripts $\alpha$
and $\beta$:
\begin{equation}
\mathcal{P}^\alpha(z) = \left[ 1 + \mathcal{P}^\beta(z)
 (\beta-\alpha) \right]^{-1} \mathcal{P}^\beta(z) , \qquad
S^\alpha = \left[ 1 + S^\beta (\beta-\alpha) \right]^{-1} S^\beta ,
\label{eq_trps}
\end{equation}
where the quantities $\alpha$ and $\beta$ in the brackets denote
nonrandom site-diagonal matrices of the screening constants.
\cite{r_1997_tdk, r_1986_apj}
Let us introduce $\eta = \beta-\alpha$ and
\begin{eqnarray}
M(z) & = & 1 + \mathcal{P}^\beta(z) \eta ,
\qquad
M^\times(z) = 1 + \eta \mathcal{P}^\beta(z) ,
\nonumber\\
K & = & 1 + S^\beta \eta , 
\qquad 
K^\times = 1 + \eta S^\beta , 
\label{eq_defmzk}
\end{eqnarray}
and let us abbreviate $\bar{g}_\pm = \lim_{\epsilon \to 0^+}
\bar{g}(E_\mathrm{F} \pm \mathrm{i} \epsilon)$ 
[and similarly for other energy dependent quantities, such as the
coherent potential functions $\mathcal{P}(z)$, the site-diagonal
matrices $M(z)$ and $M^\times(z)$, and the single-site T matrices
$t_\mathbf{R}(z)$].  
The transformation properties of the average auxiliary
Green's function $\bar{g}(z)$ can be summarized as
\begin{eqnarray}
\bar{g}^\alpha(z) & = & M^\times(z) \bar{g}^\beta(z) M(z)
 - \eta M(z) = K^\times \bar{g}^\beta(z) K + K^\times \eta 
\nonumber\\
 & = & M^\times(z) \bar{g}^\beta(z) K 
 = K^\times \bar{g}^\beta(z) M(z) ,
\label{eq_trgf}
\end{eqnarray}
the energy derivative of $\bar{g}(z)$ transforms as
\begin{equation}
\bar{g}'^{,\alpha}(z) = K^\times \left[ \bar{g}'^{,\beta}(z)
 M(z) + \bar{g}^\beta(z) M'(z) \right] 
 = K^\times \bar{g}'^{,\beta}(z) K , 
\label{eq_trgfder}
\end{equation}
and the transformation rule for the difference of two Green's
functions is
\begin{equation}
\bar{g}^\alpha_+ - \bar{g}^\alpha_-  =   
K^\times ( \bar{g}^\beta_+ - \bar{g}^\beta_- ) K . 
\label{eq_trgfdif}
\end{equation}
The transformation properties of the effective velocities
$\tilde{v}_\mu$ and $v_\nu$ are given by
\begin{equation}
\tilde{v}^\alpha_\mu = K^{-1} \tilde{v}^\beta_\mu (K^\times)^{-1} , 
\qquad
v^\alpha_\nu = K^{-1} v^\beta_\nu (K^\times)^{-1} . 
\label{eq_trvel}
\end{equation}
All these transformation rules can be proved by procedures similar to
those found in Refs.~\onlinecite{r_2012_tkd, r_1986_apj}, taking into
account also the spin independence of matrices $\alpha$, $\beta$, 
$S^\alpha$, $S^\beta$, and $K$.

The on-site blocks $M_\mathbf{R}(z)$ and $M^\times_\mathbf{R}(z)$ of
the respective site-diagonal matrices $M(z)$ and $M^\times(z)$ enter
also the transformation of the LMTO-CPA single-site T matrices
$t_\mathbf{R}(z)$, namely,
\begin{equation}
t^\beta_\mathbf{R}(z) = M_\mathbf{R}(z) t^\alpha_\mathbf{R}(z)
M^\times_\mathbf{R}(z) ,
\label{eq_trtm}
\end{equation}
see Refs.~\onlinecite{r_1997_tdk, r_2000_tkd} for more details.

\subsection{Coherent contributions\label{app_coher}}

The transformation of the coherent part of the Fermi surface term
(\ref{eq_surf}) is
\begin{equation}
\tilde{\sigma}^{(1),\alpha}_{\mu\nu,\mathrm{coh}} 
 = \sigma_0 \mathrm{Tr} \left\{ \tilde{v}^\alpha_\mu 
   ( \bar{g}^\alpha_+ - \bar{g}^\alpha_- ) v^\alpha_\nu 
     \bar{g}^\alpha_- - \tilde{v}^\alpha_\mu \bar{g}^\alpha_+ 
  v^\alpha_\nu ( \bar{g}^\alpha_+ - \bar{g}^\alpha_- ) \right\} 
 = \tilde{\sigma}^{(1),\beta}_{\mu\nu,\mathrm{coh}} + Z_{\mu\nu} ,
\label{eq_tr1coh}
\end{equation}
where the remainder can be written as
\begin{eqnarray}
Z_{\mu\nu} & = & \sigma_0 \mathrm{Tr} \left\{ 
 \tilde{v}^\beta_\mu ( \bar{g}^\beta_+ - \bar{g}^\beta_- ) 
 v^\beta_\nu \eta K^{-1} - \tilde{v}^\beta_\mu \eta K^{-1} 
 v^\beta_\nu ( \bar{g}^\beta_+ - \bar{g}^\beta_- ) \right\} 
\nonumber\\
 & = & \sigma_0 \mathrm{Tr} \left\{ 
 Y_{\mu\nu} ( \bar{g}^\beta_+ - \bar{g}^\beta_- ) \right\} ,
\qquad
Y_{\mu\nu} = v^\beta_\nu \eta K^{-1} \tilde{v}^\beta_\mu 
 - \tilde{v}^\beta_\mu \eta K^{-1} v^\beta_\nu .
\label{eq_zymunu}
\end{eqnarray}
This result proves that for metallic systems, the coherent part of 
$\tilde{\sigma}^{(1)}_{\mu\nu}$ depends on the particular LMTO
representation. 

The coherent part of the Fermi sea term (\ref{eq_seacoh})
transforms as
\begin{equation}
\tilde{\sigma}^{(2),\alpha}_{\mu\nu,\mathrm{coh}} = 
\sigma_0 \int_C \mathrm{d}z  \mathrm{Tr} 
\left\{ \tilde{v}^\alpha_\mu \bar{g}'^{,\alpha}(z) v^\alpha_\nu
 \bar{g}^\alpha(z) - \tilde{v}^\alpha_\mu \bar{g}^\alpha(z)
 v^\alpha_\nu \bar{g}'^{,\alpha}(z) \right\} =
\tilde{\sigma}^{(2),\beta}_{\mu\nu,\mathrm{coh}} + R_{\mu\nu} ,
\label{eq_tr2coh}
\end{equation}
where the remainder is
\begin{equation}
R_{\mu\nu} = \sigma_0 \int_C \mathrm{d}z \mathrm{Tr} \left\{ 
 \tilde{v}^\beta_\mu \bar{g}'^{,\beta}(z) v^\beta_\nu \eta K^{-1}
 - \tilde{v}^\beta_\mu \eta K^{-1} v^\beta_\nu \bar{g}'^{,\beta}(z) 
 \right\} .
\label{eq_rmunu}
\end{equation}
This remainder can be rewritten with the use of 
$\int_C \mathrm{d}z \bar{g}'^{,\beta}(z) =
\bar{g}^\beta_- - \bar{g}^\beta_+$, which yields
\begin{equation}
R_{\mu\nu} = \sigma_0 \mathrm{Tr} \left\{ Y_{\mu\nu} 
( \bar{g}^\beta_- - \bar{g}^\beta_+ ) \right\} = - Z_{\mu\nu} .
\label{eq_rfin}
\end{equation}
This result proves that the coherent part of
$\tilde{\sigma}^{(2)}_{\mu\nu}$ depends on the choice of the
LMTO representation as well, but the sum 
$\tilde{\sigma}^{(1)}_{\mu\nu,\mathrm{coh}} + 
\tilde{\sigma}^{(2)}_{\mu\nu,\mathrm{coh}}$, i.e., the total
coherent part of $\tilde{\sigma}_{\mu\nu}$, is strictly invariant,
as mentioned in Section \ref{ss_tpsdct}.

\subsection{Incoherent part of the Fermi surface
            term \label{app_vcsurf}}

For the vertex corrections to the Fermi surface term (\ref{eq_surf}),
transformation properties are needed for all quantities entering the
general expression for the LMTO vertex corrections \cite{r_2006_ctk}.
The transformation rule for a two-particle quantity 
$\chi^{\Lambda_1 \Lambda_2}_{\mathbf{R}_1 \mathbf{R}_2}$ depending
only on elements of the average auxiliary Green's functions between
different sites ($\mathbf{R}_1 \ne \mathbf{R}_2$) and defined as
\begin{equation}
\chi^{\Lambda_1 \Lambda_2}_{\mathbf{R}_1 \mathbf{R}_2}
= ( 1 - \delta_{\mathbf{R}_1 \mathbf{R}_2} ) 
 (\bar{g}_+)^{L_1 L_2}_{\mathbf{R}_1 \mathbf{R}_2} 
  (\bar{g}_-)^{L'_2 L'_1}_{\mathbf{R}_2 \mathbf{R}_1} ,
\label{eq_chidef}
\end{equation}
where $\Lambda_1 = (L_1, L'_1)$, $\Lambda_2 = (L_2, L'_2)$,  
is given with the help of Eq.~(\ref{eq_trgf}) by
\begin{equation}
\chi^\alpha = \Pi \chi^\beta \tilde{\Pi} ,
\label{eq_trchi}
\end{equation}
where we introduced site-diagonal quantities
$\Pi^{\Lambda_1 \Lambda_2}_{\mathbf{R}_1 \mathbf{R}_2} = 
\delta_{\mathbf{R}_1 \mathbf{R}_2} 
\Pi^{\Lambda_1 \Lambda_2}_{\mathbf{R}_1}$ and 
$\tilde{\Pi}^{\Lambda_1 \Lambda_2}_{\mathbf{R}_1 \mathbf{R}_2} 
 = \delta_{\mathbf{R}_1 \mathbf{R}_2}
 \tilde{\Pi}^{\Lambda_1 \Lambda_2}_{\mathbf{R}_1}$, where
\begin{equation}
\Pi^{\Lambda_1 \Lambda_2}_\mathbf{R} = 
\left( M^\times_{+,\mathbf{R}} \right)^{L_1 L_2}
\left( M_{-,\mathbf{R}} \right)^{L'_2 L'_1} , 
\qquad 
\tilde{\Pi}^{\Lambda_1 \Lambda_2}_\mathbf{R} = 
\left( M_{+,\mathbf{R}}  \right)^{L_1 L_2}
\left( M^\times_{-,\mathbf{R}} \right)^{L'_2 L'_1} .
\label{eq_pitipi} 
\end{equation}
The site-diagonal quantity 
$w^{\Lambda_1 \Lambda_2}_{\mathbf{R}_1 \mathbf{R}_2} = 
\delta_{\mathbf{R}_1 \mathbf{R}_2} 
w^{\Lambda_1 \Lambda_2}_{\mathbf{R}_1}$, where
$w^{\Lambda_1 \Lambda_2}_\mathbf{R} = \left\langle 
t^{L_1 L_2}_{+,\mathbf{R}} t^{L'_2 L'_1}_{-,\mathbf{R}}
\right\rangle$, satisfies the transformation relation
\begin{equation}
w^\beta = \tilde{\Pi} w^\alpha \Pi ,
\label{eq_trw}
\end{equation}
which follows from Eq.~(\ref{eq_trtm}).
As a consequence of the rules (\ref{eq_trchi}) and (\ref{eq_trw}),
the matrix $\Delta = w^{-1} - \chi$ and its inverse transform as
\begin{equation}
\Delta^\alpha = \Pi \Delta^\beta \tilde{\Pi} , \qquad
(\Delta^\alpha)^{-1} = \tilde{\Pi}^{-1} 
(\Delta^\beta)^{-1} \Pi^{-1} .
\label{eq_trdelin}
\end{equation}
For transformations of the on-site blocks $(\bar{g}_+ 
v_\nu \bar{g}_-)^{L_1 L'_1}_{\mathbf{R} \mathbf{R}} \equiv
(\bar{g}_+ v_\nu \bar{g}_-)^{\Lambda_1}_\mathbf{R}$ and $(\bar{g}_- 
\tilde{v}_\mu \bar{g}_+)^{L'_1 L_1}_{\mathbf{R} \mathbf{R}} \equiv 
(\bar{g}_- \tilde{v}_\mu \bar{g}_+)^{\tilde{\Lambda}_1}_\mathbf{R}$,
one can use relations (\ref{eq_trgf}) and (\ref{eq_trvel}) for
the Green's functions and velocities, respectively.
The result is
\begin{equation}
(\bar{g}^\alpha_+ v^\alpha_\nu 
\bar{g}^\alpha_-)^{\Lambda_1}_\mathbf{R}
= \sum_{\Lambda_2} \Pi^{\Lambda_1 \Lambda_2}_\mathbf{R} 
(\bar{g}^\beta_+ v^\beta_\nu 
\bar{g}^\beta_-)^{\Lambda_2}_\mathbf{R} ,
\qquad
(\bar{g}^\alpha_- \tilde{v}^\alpha_\mu 
\bar{g}^\alpha_+)^{\tilde{\Lambda}_1}_\mathbf{R}
= \sum_{\Lambda_2} (\bar{g}^\beta_- \tilde{v}^\beta_\mu 
\bar{g}^\beta_+)^{\tilde{\Lambda}_2}_\mathbf{R} 
\tilde{\Pi}^{\Lambda_2 \Lambda_1}_\mathbf{R} , 
\label{eq_trgvg}
\end{equation}
where the symbols $\tilde{\Lambda}_1 = (L'_1, L_1)$
and $\tilde{\Lambda}_2 = (L'_2, L_2)$ denote indices transposed 
to $\Lambda_1 = (L_1, L'_1)$ and $\Lambda_2 = (L_2, L'_2)$, 
respectively. 

The transformation of the vertex corrections to the Fermi surface
term (\ref{eq_surf}) is now straightforward \cite{r_2006_ctk}.
The identity (\ref{eq_gvgon}) yields
$\mathrm{Tr} \left\langle  \tilde{v}_\mu g_+ 
v_\nu g_+ \right\rangle_\mathrm{VC} =
\mathrm{Tr} \left\langle  \tilde{v}_\mu g_- 
v_\nu g_- \right\rangle_\mathrm{VC} = 0$, 
so that 
$\tilde{\sigma}^{(1)}_{\mu\nu,\mathrm{VC}} = 
2 \sigma_0 \mathrm{Tr} \left\langle  \tilde{v}_\mu g_+ 
v_\nu g_- \right\rangle_\mathrm{VC}$ and
\begin{equation}
\tilde{\sigma}^{(1),\alpha}_{\mu\nu,\mathrm{VC}} = 2 \sigma_0 
\sum_{\mathbf{R}_1 \Lambda_1} \sum_{\mathbf{R}_2 \Lambda_2} 
(\bar{g}^\alpha_- \tilde{v}^\alpha_\mu 
\bar{g}^\alpha_+)^{\tilde{\Lambda}_1}_{\mathbf{R}_1}
\left[ (\Delta^\alpha)^{-1} 
\right]^{\Lambda_1 \Lambda_2}_{\mathbf{R}_1 \mathbf{R}_2}
(\bar{g}^\alpha_+ v^\alpha_\nu 
\bar{g}^\alpha_-)^{\Lambda_2}_{\mathbf{R}_2} .
\label{eq_sig1vc}
\end{equation}
The last relation combined with the transformations
(\ref{eq_trdelin}) and (\ref{eq_trgvg}) leads to the invariance
of the vertex corrections to the Fermi surface term,
$\tilde{\sigma}^{(1),\alpha}_{\mu\nu,\mathrm{VC}}
= \tilde{\sigma}^{(1),\beta}_{\mu\nu,\mathrm{VC}}$.

\subsection{Incoherent part of the Fermi sea term \label{app_vcsea}}

The vertex corrections to the Fermi sea term (\ref{eq_seavc}) can
be written as $(-2 \sigma_0) \int_C \mathrm{d}z Q(z)$, where the
quantity $Q(z)$ in the LMTO representation $\alpha$ is given
explicitly by
\begin{equation}
Q^\alpha = \sum_{\mathbf{R}_1 \Lambda_1} 
\sum_{\mathbf{R}_2 \Lambda_2} (\bar{g}^\alpha \tilde{v}^\alpha_\mu 
\bar{g}^\alpha)^{\tilde{\Lambda}_1}_{\mathbf{R}_1}
\left[ (\Delta^\alpha)^{-1} 
\right]^{\Lambda_1 \Lambda_2}_{\mathbf{R}_1 \mathbf{R}_2}
(\bar{g}^\alpha v^\alpha_\nu \bar{g}'^{,\alpha}
 )^{\Lambda_2}_{\mathbf{R}_2} ,
\label{eq_qz}
\end{equation}
where all energy arguments (equal to $z$) have been omitted,
see Eq.~(\ref{eq_seavc}), where $\Lambda_1 = (L_1, L'_1)$,
$\Lambda_2 = (L_2, L'_2)$, $\tilde{\Lambda}_1 = (L'_1, L_1)$,
and $\tilde{\Lambda}_2 = (L'_2, L_2)$, and where the same
identifications have been done as before, namely,
$(\bar{g} v_\nu \bar{g}')^{L_1 L'_1}_{\mathbf{R} \mathbf{R}} \equiv
(\bar{g} v_\nu \bar{g}')^{\Lambda_1}_\mathbf{R}$ and $(\bar{g} 
\tilde{v}_\mu \bar{g})^{L'_1 L_1}_{\mathbf{R} \mathbf{R}} \equiv 
(\bar{g} \tilde{v}_\mu \bar{g})^{\tilde{\Lambda}_1}_\mathbf{R}$.

The transformation of individual factors in Eq.~(\ref{eq_qz}) is
similar to the previous case of the Fermi surface term.
In particular, the transformation rules for $\Delta^\alpha$ and
its inverse, Eq.~(\ref{eq_trdelin}), remain valid but with the
quantities $\Pi_\mathbf{R}$ and $\tilde{\Pi}_\mathbf{R}$
defined as
\begin{equation}
\Pi^{\Lambda_1 \Lambda_2}_\mathbf{R} = 
\left( M^\times_\mathbf{R} \right)^{L_1 L_2}
\left( M_\mathbf{R} \right)^{L'_2 L'_1} , 
\qquad 
\tilde{\Pi}^{\Lambda_1 \Lambda_2}_\mathbf{R} = 
\left( M_\mathbf{R}  \right)^{L_1 L_2}
\left( M^\times_\mathbf{R} \right)^{L'_2 L'_1} ,
\label{eq_pitipi2} 
\end{equation}
where the matrices $M_\mathbf{R}$ and $M^\times_\mathbf{R}$
are taken with the same (omitted) energy argument $z$.
The transformation of $(\bar{g} \tilde{v}_\mu 
\bar{g})^{\tilde{\Lambda}_1}_\mathbf{R}$ is similar to
Eq.~(\ref{eq_trgvg}), namely,
\begin{equation}
(\bar{g}^\alpha \tilde{v}^\alpha_\mu 
\bar{g}^\alpha)^{\tilde{\Lambda}_1}_\mathbf{R}
= \sum_{\Lambda_2} (\bar{g}^\beta \tilde{v}^\beta_\mu 
\bar{g}^\beta)^{\tilde{\Lambda}_2}_\mathbf{R} 
\tilde{\Pi}^{\Lambda_2 \Lambda_1}_\mathbf{R} , 
\label{eq_trgvg2}
\end{equation}
while the transformation of
$(\bar{g} v_\nu \bar{g}')^{\Lambda_1}_\mathbf{R}$ is based on
relation~(\ref{eq_trgfder}), which leads to
\begin{equation}
(\bar{g}^\alpha v^\alpha_\nu 
\bar{g}'^{,\alpha})^{L_1 L'_1}_{\mathbf{R} \mathbf{R}} =
(M^\times \bar{g}^\beta v^\beta_\nu \bar{g}'^{,\beta}
 M)^{L_1 L'_1}_{\mathbf{R} \mathbf{R}} +
(M^\times \bar{g}^\beta v^\beta_\nu \bar{g}^\beta
 M')^{L_1 L'_1}_{\mathbf{R} \mathbf{R}} ,
\label{eq_trgvgpaux}
\end{equation}
where the second term on the r.h.s.\ vanishes due to the
identity~(\ref{eq_gvgon}) and due to the site-diagonal nature
of matrices $M$, $M^\times$ and $M'$. 
This yields  
\begin{equation}
(\bar{g}^\alpha v^\alpha_\nu \bar{g}'^{,\alpha}
 )^{\Lambda_1}_\mathbf{R}
= \sum_{\Lambda_2} \Pi^{\Lambda_1 \Lambda_2}_\mathbf{R} 
(\bar{g}^\beta v^\beta_\nu \bar{g}'^{,\beta}
 )^{\Lambda_2}_\mathbf{R} ,
\label{eq_trgvgp}
\end{equation}
again in analogy with Eq.~(\ref{eq_trgvg}).
The use of the rules (\ref{eq_trdelin}), (\ref{eq_trgvg2}) and
(\ref{eq_trgvgp}) in Eq.~(\ref{eq_qz}) leads to $Q^\alpha(z) =
Q^\beta(z)$, which proves the transformation invariance of the
vertex corrections to the Fermi sea term, 
$\tilde{\sigma}^{(2),\alpha}_{\mu\nu,\mathrm{VC}}
= \tilde{\sigma}^{(2),\beta}_{\mu\nu,\mathrm{VC}}$.

This completes the proof of the invariance of the total
spin-dependent conductivity tensor $\tilde{\sigma}_{\mu\nu}$
(\ref{eq_sigma}) with respect to the choice of the LMTO
representation.

\section{(Supplemental material) Equivalence of the KKR and
         TB-LMTO conductivity tensors}

In this part, we sketch a proof of equivalence of the conductivity
tensors based (i) on the usual, continuous coordinates used
in the KKR method, and (ii) on the modified, piecewise constant
coordinates employed in the TB-LMTO technique. \cite{r_2002_tkd}
In both cases we consider the same one-particle Hamiltonian $H$ and
denote its retarded and advanced Green's functions as
$G_\pm(E) = (E \pm i 0 - H)^{-1}$, where $E$ is a real energy
variable.

\bigskip

\paragraph{Basic definitions and relations.}

Let us start with a few basic definitions and relations.
If $A$ denotes a Hermitian operator representing a one-particle
observable quantity, its thermodynamic equilibrium average value
$Q(A)$ can be expressed as
\begin{equation}
Q(A) = \frac{i}{2\pi} \int \mathrm{d}E f(E) \mathrm{Tr}
\{ A D(E) \} ,
\label{eqsm_qdef}
\end{equation}
where we abbreviated
\begin{equation}
D(E) = G_+(E) - G_-(E) 
\label{eqsm_ddef}
\end{equation}
and where $f(E)$ denotes the Fermi-Dirac function.
Let $B$ denote another one-particle Hermitian operator which, added
as a small perturbation to the unperturbed Hamiltonian $H$, defines
a new thermodynamic equilibrium (with the same Fermi-Dirac function)
of the system and let us assume that this perturbation leaves the
operator $A$ unchanged.
The corresponding first-order changes in the Green's functions
are $\delta G_\pm (E) = G_\pm (E) B G_\pm (E)$ which induces
a change in the average value of $A$ equal to $\delta Q(A) = R(A,B)$,
where
\begin{equation}
R(A,B) = \frac{i}{2\pi} \int \mathrm{d}E f(E) \mathrm{Tr}
\{ A G_+(E) B G_+(E) - A G_-(E) B G_-(E) \} 
\label{eqsm_rdef}
\end{equation}
defines the relevant isothermic linear-response coefficient
(susceptibility).
An obvious symmetry relation
\begin{equation}
R(A,B) = R(B,A)
\label{eqsm_rsym}
\end{equation}
is valid owing to the cyclic property of the trace. 
For the adiabatic linear response, we introduce a symbol
\begin{equation}
C(A_1 , A_2) = \frac{1}{2\pi} \int \mathrm{d}E f(E)
 \mathrm{Tr} \{ A_1 D(E) A_2 G^2_-(E) 
 - A_1 G^2_+(E) A_2 D(E) \} , 
\label{eqsm_cdef}
\end{equation}
where $A_1$ and $A_2$ are Hermitian operators.
If these operators are identified with two components of the velocity
(current) operator, then $C(A_1 , A_2)$ defines (apart from a
prefactor) an element of the Kubo conductivity tensor as given by the
formula of Bastin et al. \cite{r_1971_blb}
Note that for energy derivative of the Green's functions, the
well-known relation $G'_\pm (E) = - G^2_\pm (E)$ has been used.

\bigskip

\paragraph{Further relations.}

We prove now several useful relations among the introduced
quantities $Q(A)$, $R(A,B)$ and $C(A_1 , A_2)$.
Let us assume that the operator $A_2$ is a time derivative of 
an operator $B_2$, so that ($\hbar = 1$) 
\begin{equation}
A_2 = -i [ B_2 , H ] , 
\label{eqsm_a2b2h}
\end{equation}
and that $B_2$ can be considered as a perturbation with the property
mentioned before Eq.~(\ref{eqsm_rdef}).
Then one can write 
\begin{equation}
A_2 = i B_2 (E-H) - i (E-H) B_2
\label{eqsm_a2b2emh}
\end{equation}
in Eq.~(\ref{eqsm_cdef}) and one can use relations
\begin{equation}
(E-H) G_\pm(E) = G_\pm(E) (E-H) = 1
\label{eqsm_wg}
\end{equation}
and
\begin{equation}
(E-H) D(E) = D(E) (E-H) = 0 ,
\label{eqsm_wd}
\end{equation}
which follow from the definitions of the Green's functions
and from Eq.~(\ref{eqsm_ddef}).
The trace in Eq.~(\ref{eqsm_cdef}) can be then simplified as
\begin{eqnarray}
 & & \mathrm{Tr} \{ A_1 D(E) A_2 G^2_-(E) 
                  - A_1 G^2_+(E) A_2 D(E) \} 
\nonumber\\
 & = & i\, \mathrm{Tr} \{ A_1 D(E) B_2 (E-H) G^2_-(E)
                        - A_1 D(E) (E-H) B_2 G^2_-(E)   
\nonumber\\
 & & \quad {} - A_1 G^2_+(E) B_2 (E-H) D(E)
              + A_1 G^2_+(E) (E-H) B_2 D(E) \}
\nonumber\\
 & = & i\, \mathrm{Tr} \{ A_1 D(E) B_2 G_-(E)
                      + A_1 G_+(E) B_2 D(E) \}
\nonumber\\
 & = & i\, \mathrm{Tr} \{ A_1 G_+(E) B_2 G_+(E)
                        - A_1 G_-(E) B_2 G_-(E) \} ,
\label{eqsm_wa}
\end{eqnarray}
which yields a relation
\begin{equation}
C(A_1 , A_2) = R(A_1 , B_2) .
\label{eqsm_cr1}
\end{equation}
Similarly, if $A_1$ in Eq.~(\ref{eqsm_cdef}) is a time derivative
of an operator $B_1$ with the property mentioned before
Eq.~(\ref{eqsm_rdef}),
\begin{equation}
A_1 = -i [ B_1 , H ] ,
\label{eqsm_a1b1h}
\end{equation}
one can prove that
\begin{equation}
C(A_1 , A_2) = - R(B_1 , A_2) .
\label{eqsm_cr2}
\end{equation}
Finally, one can prove in a similar way that if both operators $A_1$
and $A_2$ in Eq.~(\ref{eqsm_cdef}) can be written according to
Eq.~(\ref{eqsm_a1b1h}) and Eq.~(\ref{eqsm_a2b2h}), respectively, then
\begin{equation}
C(A_1 , A_2) =  Q ( i [ B_1 , B_2 ] ) .
\label{eqsm_cq}
\end{equation}
The proof of equivalence of both conductivity tensors employs
the relations (\ref{eqsm_cr1}, \ref{eqsm_cr2}, \ref{eqsm_cq}).

\bigskip

\paragraph{Proof of equivalence.}

We denote the operator of the usual, continuous coordinate as
$r_\mu$ ($\mu = x, y, z$) and that of the modified coordinate,
constant in each Wigner-Seitz cell, as $r^0_\mu$; their difference
is denoted as $\xi_\mu$, so that
\begin{equation}
r_\mu = r^0_\mu + \xi_\mu .
\label{eqsm_coord}
\end{equation}
The time derivatives of the coordinate operators $r_\mu$, $r^0_\mu$
and $\xi_\mu$ due to the Hamiltonian $H$ are denoted as
$V_\mu$, $V^0_\mu$ and $U_\mu$, respectively.
These velocity operators satisfy the obvious relation
\begin{equation}
V_\mu = V^0_\mu + U_\mu .
\label{eqsm_veloc}
\end{equation}
It should be noted that only the difference coordinate $\xi_\mu$,
which is bounded due to a finite size of the Wigner-Seitz cells,
can be used as a perturbation that defines a new thermodynamic
equilibrium; the other two coordinates ($r_\mu$, $r^0_\mu$), which
are unbounded in infinite solids, do not possess this property.
This means that only the relation
\begin{equation}
U_\mu = -i [ \xi_\mu , H ]
\label{eqsm_udef}
\end{equation}
can be used as a special case of relations (\ref{eqsm_a2b2h})
and (\ref{eqsm_a1b1h}).

The elements of the KKR conductivity tensor can be (apart from a
prefactor) identified with $C(V_\mu , V_\nu)$ while their TB-LMTO
counterparts are given by $C(V^0_\mu , V^0_\nu)$.
We have then 
\begin{equation}
C(V_\mu , V_\nu) = C(V^0_\mu , V^0_\nu) + C(V^0_\mu , U_\nu)
 + C(U_\mu , V^0_\nu) + C(U_\mu , U_\nu)
\label{eqsm_csum}
\end{equation}
as a simple consequence of Eq.~(\ref{eqsm_cdef}) and
Eq.~(\ref{eqsm_veloc}). 
The last term in Eq.~(\ref{eqsm_csum}) vanishes due to
Eq.~(\ref{eqsm_udef}) and the rule (\ref{eqsm_cq}), 
\begin{equation}
C(U_\mu , U_\nu) = Q( i [ \xi_\mu , \xi_\nu ] ) = 0 ,
\label{eqsm_cuu}
\end{equation}
since $[ \xi_\mu , \xi_\nu ] = 0$ (the coordinate operators commute
mutually).
Before the treatment of the other terms in Eq.~(\ref{eqsm_csum}),
let us discuss the quantity $C(V_\mu , U_\nu)$.
It holds
\begin{equation}
C(V_\mu , U_\nu) = R(V_\mu , \xi_\nu )
\label{eqsm_cvua}
\end{equation}
as a consequence of Eq.~(\ref{eqsm_udef}) and Eq.~(\ref{eqsm_cr1}).
Let us prove that the r.h.s.\ of Eq.~(\ref{eqsm_cvua}) vanishes.
We start with a general relation
\begin{equation}
Q(V_\mu) = 0 ,
\label{eqsm_qv}
\end{equation}
which is nothing but an exact vanishing of the average value of the
total electric current in thermodynamic equilibrium. 
If we add the operator $\xi_\nu$ as a perturbation to the Hamiltonian
$H$ (which defines a new equilibrium of the system) and if we take
into account that this perturbation does not modify the velocity
operator $V_\mu$ (since $[ r_\mu , \xi_\nu ] = 0$),
we obtain the same exact vanishing of the average value of the total
electric current also for the perturbed system.
This means that the corresponding linear-response coefficient
(\ref{eqsm_rdef}) vanishes as well, hence
\begin{equation}
R(V_\mu , \xi_\nu) = 0 .
\label{eqsm_rvxi}
\end{equation}
Combination of this result with Eq.~(\ref{eqsm_cvua}) and
Eq.~(\ref{eqsm_veloc}) yields
\begin{equation}
C(V_\mu , U_\nu) = C(V^0_\mu , U_\nu) + C(U_\mu , U_\nu) = 0 .
\label{eqsm_cvuz}
\end{equation}
By using the previous relation (\ref{eqsm_cuu}), we get
\begin{equation}
C(V^0_\mu , U_\nu) = 0 ,
\label{eqsm_cv0u}
\end{equation}
so that the second term on the r.h.s.\ of Eq.~(\ref{eqsm_csum})
vanishes.
Let us consider finally the term $C(U_\mu , V^0_\nu)$.
With application of Eq.~(\ref{eqsm_udef}), Eq.~(\ref{eqsm_cr2}), 
Eq.~(\ref{eqsm_rsym}), Eq.~(\ref{eqsm_cr1}), and
Eq.~(\ref{eqsm_cv0u}), we get
\begin{equation}
 C(U_\mu , V^0_\nu) = - R(\xi_\mu , V^0_\nu) =
 - R(V^0_\nu , \xi_\mu) = - C(V^0_\nu , U_\mu) = 0 .
\label{eqsm_cuv0}
\end{equation}
The last three terms on the r.h.s.\ of Eq.~(\ref{eqsm_csum}) thus
vanish, see Eqs.~(\ref{eqsm_cuu}, \ref{eqsm_cv0u}, \ref{eqsm_cuv0}),
which yields
\begin{equation}
C(V_\mu , V_\nu) = C(V^0_\mu , V^0_\nu) .
\label{eqsm_equiv}
\end{equation}
This completes the proof of equivalence of both conductivity
tensors.

%\bibliography{b_}

\providecommand{\noopsort}[1]{}\providecommand{\singleletter}[1]{#1}%

\end{document}